\documentclass[aps,preprint,superscriptaddress, pra]{revtex4-2}

\usepackage{graphicx}
\usepackage{amsmath}
\usepackage{color}
\usepackage{bm}
\usepackage{xcolor}
\usepackage{hyperref}
\hypersetup{
    colorlinks,
    linkcolor={red!50!black},
    citecolor={blue!50!black},
    urlcolor={blue!80!black}
  }
\usepackage{braket}
\usepackage{multirow}
\usepackage{datatool, filecontents}
\DTLsetseparator{ = }
\DTLloaddb[noheader, keys={key,value}]{dataA}{capRateListAvg_doublet_1B_FSI_N2LOsim_450.dat}
\DTLloaddb[noheader, keys={key,value}]{dataB}{capRateListAvg_doublet_1B_FSI_N2LOsim_475.dat}
\DTLloaddb[noheader, keys={key,value}]{dataC}{capRateListAvg_doublet_1B_FSI_N2LOsim_500.dat}
\DTLloaddb[noheader, keys={key,value}]{dataD}{capRateListAvg_doublet_1B_FSI_N2LOsim_525.dat}
\DTLloaddb[noheader, keys={key,value}]{dataE}{capRateListAvg_doublet_1B_FSI_N2LOsim_550.dat}
\DTLloaddb[noheader, keys={key,value}]{dataF}{capRateListAvg_doublet_1B_FSI_N2LOsim_575.dat}
\DTLloaddb[noheader, keys={key,value}]{dataG}{capRateListAvg_doublet_1B_FSI_N2LOsim_600.dat}

\DTLloaddb[noheader, keys={key,value}]{dataH}{capRateListAvg_quartet_1B_FSI_N2LOsim_450.dat}
\DTLloaddb[noheader, keys={key,value}]{dataI}{capRateListAvg_quartet_1B_FSI_N2LOsim_475.dat}
\DTLloaddb[noheader, keys={key,value}]{dataJ}{capRateListAvg_quartet_1B_FSI_N2LOsim_500.dat}
\DTLloaddb[noheader, keys={key,value}]{dataK}{capRateListAvg_quartet_1B_FSI_N2LOsim_525.dat}
\DTLloaddb[noheader, keys={key,value}]{dataL}{capRateListAvg_quartet_1B_FSI_N2LOsim_550.dat}
\DTLloaddb[noheader, keys={key,value}]{dataM}{capRateListAvg_quartet_1B_FSI_N2LOsim_575.dat}
\DTLloaddb[noheader, keys={key,value}]{dataN}{capRateListAvg_quartet_1B_FSI_N2LOsim_600.dat}

\DTLloaddb[noheader, keys={key,value}]{dataAA}{capRateListAvg_doublet_1B2B_FSI_N2LOsim_450.dat}
\DTLloaddb[noheader, keys={key,value}]{dataBB}{capRateListAvg_doublet_1B2B_FSI_N2LOsim_500.dat}
\DTLloaddb[noheader, keys={key,value}]{dataCC}{capRateListAvg_doublet_1B2B_FSI_N2LOsim_550.dat}
\DTLloaddb[noheader, keys={key,value}]{dataDD}{capRateListAvg_doublet_1B2B_FSI_N2LOsim_600.dat}

\DTLloaddb[noheader, keys={key,value}]{dataEE}{capRateListAvg_quartet_1B2B_FSI_N2LOsim_450.dat}
\DTLloaddb[noheader, keys={key,value}]{dataFF}{capRateListAvg_quartet_1B2B_FSI_N2LOsim_500.dat}
\DTLloaddb[noheader, keys={key,value}]{dataGG}{capRateListAvg_quartet_1B2B_FSI_N2LOsim_550.dat}
\DTLloaddb[noheader, keys={key,value}]{dataHH}{capRateListAvg_quartet_1B2B_FSI_N2LOsim_600.dat}

\DTLloaddb[noheader, keys={key,value}]{dataDoubletLO}{capRateList_LO_doublet_1B2B_FSI_CHP_NNLO.dat}
\DTLloaddb[noheader, keys={key,value}]{dataDoubletNLO}{capRateList_NLO_doublet_1B2B_FSI_CHP_NNLO.dat}
\DTLloaddb[noheader, keys={key,value}]{dataDoubletN2LO}{capRateList_N2LO_doublet_1B2B_FSI_CHP_NNLO.dat}

\DTLloaddb[noheader, keys={key,value}]{dataQuartetLO}{capRateList_LO_quartet_1B2B_FSI_CHP_NNLO.dat}
\DTLloaddb[noheader, keys={key,value}]{dataQuartetNLO}{capRateList_NLO_quartet_1B2B_FSI_CHP_NNLO.dat}
\DTLloaddb[noheader, keys={key,value}]{dataQuartetN2LO}{capRateList_N2LO_quartet_1B2B_FSI_CHP_NNLO.dat}

\newcommand{\potA}[1]{\DTLfetch{dataA}{key}{#1}{value}}
\newcommand{\potB}[1]{\DTLfetch{dataB}{key}{#1}{value}}
\newcommand{\potC}[1]{\DTLfetch{dataC}{key}{#1}{value}}
\newcommand{\potD}[1]{\DTLfetch{dataD}{key}{#1}{value}}
\newcommand{\potE}[1]{\DTLfetch{dataE}{key}{#1}{value}}
\newcommand{\potF}[1]{\DTLfetch{dataF}{key}{#1}{value}}
\newcommand{\potG}[1]{\DTLfetch{dataG}{key}{#1}{value}}

\newcommand{\potH}[1]{\DTLfetch{dataH}{key}{#1}{value}}
\newcommand{\potI}[1]{\DTLfetch{dataI}{key}{#1}{value}}
\newcommand{\potJ}[1]{\DTLfetch{dataJ}{key}{#1}{value}}
\newcommand{\potK}[1]{\DTLfetch{dataK}{key}{#1}{value}}
\newcommand{\potL}[1]{\DTLfetch{dataL}{key}{#1}{value}}
\newcommand{\potM}[1]{\DTLfetch{dataM}{key}{#1}{value}}
\newcommand{\potN}[1]{\DTLfetch{dataN}{key}{#1}{value}}

\newcommand{\potAA}[1]{\DTLfetch{dataAA}{key}{#1}{value}}
\newcommand{\potBB}[1]{\DTLfetch{dataBB}{key}{#1}{value}}
\newcommand{\potCC}[1]{\DTLfetch{dataCC}{key}{#1}{value}}
\newcommand{\potDD}[1]{\DTLfetch{dataDD}{key}{#1}{value}}

\newcommand{\potEE}[1]{\DTLfetch{dataEE}{key}{#1}{value}}
\newcommand{\potFF}[1]{\DTLfetch{dataFF}{key}{#1}{value}}
\newcommand{\potGG}[1]{\DTLfetch{dataGG}{key}{#1}{value}}
\newcommand{\potHH}[1]{\DTLfetch{dataHH}{key}{#1}{value}}

\newcommand{\potdLO}[1]{\DTLfetch{dataDoubletLO}{key}{#1}{value}}
\newcommand{\potdNLO}[1]{\DTLfetch{dataDoubletNLO}{key}{#1}{value}}
\newcommand{\potdNNLO}[1]{\DTLfetch{dataDoubletN2LO}{key}{#1}{value}}

\newcommand{\potqLO}[1]{\DTLfetch{dataQuartetLO}{key}{#1}{value}}
\newcommand{\potqNLO}[1]{\DTLfetch{dataQuartetNLO}{key}{#1}{value}}
\newcommand{\potqNNLO}[1]{\DTLfetch{dataQuartetN2LO}{key}{#1}{value}}

\newcommand{\q}{\textbf{\emph{q}}}
\newcommand{\p}{\textbf{\emph{p}}}
\newcommand{\ki}{\textbf{\emph{k}}}
\newcommand{\sa}{\pmb{\sigma}_{1}}
\newcommand{\sbb}{\pmb{\sigma}_{2}}
\newcommand{\ta}{\pmb{\tau}_{1}}
\newcommand{\tb}{\pmb{\tau}_{2}}
\newcommand{\mpi}{\emph{$m_{\pi}$}}

\def\beq{\begin{equation}}
\def\eeq{\end{equation}}

\def\bdm{\begin{displaymath}}
\def\edm{\end{displaymath}}

\def\bea{\begin{eqnarray}}
\def\eea{\end{eqnarray}}

\begin{document}
\begin{filecontents*}{capRateListAvg_doublet_1B_FSI_N2LOsim_450.dat}
CR_1S0 = 248.73
CR_3P0 = 14.87
CR_3P1 = 42.37
CR_3P2 = 71.54
CR_1D2 = 7.79
CR_3F2 = 0.97
CR_upto_J1 = 305.97
CR_sum = 386.27
\end{filecontents*}

\begin{filecontents*}{capRateListAvg_doublet_1B_FSI_N2LOsim_475.dat}
CR_1S0 = 248.14
CR_3P0 = 15.45
CR_3P1 = 42.36
CR_3P2 = 71.54
CR_1D2 = 7.79
CR_3F2 = 0.97
CR_upto_J1 = 305.95
CR_sum = 386.25
\end{filecontents*}

\begin{filecontents*}{capRateListAvg_doublet_1B_FSI_N2LOsim_500.dat}
CR_1S0 = 247.61
CR_3P0 = 15.83
CR_3P1 = 42.36
CR_3P2 = 71.55
CR_1D2 = 7.80
CR_3F2 = 0.97
CR_upto_J1 = 305.80
CR_sum = 386.12
\end{filecontents*}

\begin{filecontents*}{capRateListAvg_doublet_1B_FSI_N2LOsim_525.dat}
CR_1S0 = 247.14
CR_3P0 = 16.07
CR_3P1 = 42.36
CR_3P2 = 71.56
CR_1D2 = 7.80
CR_3F2 = 0.97
CR_upto_J1 = 305.57
CR_sum = 385.90
\end{filecontents*}

\begin{filecontents*}{capRateListAvg_doublet_1B_FSI_N2LOsim_550.dat}
CR_1S0 = 246.73
CR_3P0 = 16.24
CR_3P1 = 42.37
CR_3P2 = 71.57
CR_1D2 = 7.80
CR_3F2 = 0.97
CR_upto_J1 = 305.34
CR_sum = 385.68
\end{filecontents*}

\begin{filecontents*}{capRateListAvg_doublet_1B_FSI_N2LOsim_575.dat}
CR_1S0 = 246.37
CR_3P0 = 16.35
CR_3P1 = 42.38
CR_3P2 = 71.59
CR_1D2 = 7.80
CR_3F2 = 0.98
CR_upto_J1 = 305.10
CR_sum = 385.47
\end{filecontents*}

\begin{filecontents*}{capRateListAvg_doublet_1B_FSI_N2LOsim_600.dat}
CR_1S0 = 246.07
CR_3P0 = 16.44
CR_3P1 = 42.39
CR_3P2 = 71.60
CR_1D2 = 7.80
CR_3F2 = 0.98
CR_upto_J1 = 304.90
CR_sum = 385.28
\end{filecontents*}

\begin{filecontents*}{capRateListAvg_quartet_1B_FSI_N2LOsim_450.dat}
CR_1S0 = 6.64
CR_3P0 = 0.53
CR_3P1 = 0.49
CR_3P2 = 1.43
CR_1D2 = 1.87
CR_3F2 = 0.27
CR_upto_J1 = 7.66
CR_sum = 11.23
\end{filecontents*}

\begin{filecontents*}{capRateListAvg_quartet_1B_FSI_N2LOsim_475.dat}
CR_1S0 = 6.62
CR_3P0 = 0.55
CR_3P1 = 0.49
CR_3P2 = 1.43
CR_1D2 = 1.90
CR_3F2 = 0.27
CR_upto_J1 = 7.66
CR_sum = 11.26
\end{filecontents*}

\begin{filecontents*}{capRateListAvg_quartet_1B_FSI_N2LOsim_500.dat}
CR_1S0 = 6.61
CR_3P0 = 0.56
CR_3P1 = 0.49
CR_3P2 = 1.43
CR_1D2 = 1.93
CR_3F2 = 0.27
CR_upto_J1 = 7.66
CR_sum = 11.29
\end{filecontents*}

\begin{filecontents*}{capRateListAvg_quartet_1B_FSI_N2LOsim_525.dat}
CR_1S0 = 6.60
CR_3P0 = 0.57
CR_3P1 = 0.49
CR_3P2 = 1.44
CR_1D2 = 1.94
CR_3F2 = 0.27
CR_upto_J1 = 7.66
CR_sum = 11.31
\end{filecontents*}

\begin{filecontents*}{capRateListAvg_quartet_1B_FSI_N2LOsim_550.dat}
CR_1S0 = 6.59
CR_3P0 = 0.57
CR_3P1 = 0.49
CR_3P2 = 1.44
CR_1D2 = 1.96
CR_3F2 = 0.27
CR_upto_J1 = 7.65
CR_sum = 11.32
\end{filecontents*}

\begin{filecontents*}{capRateListAvg_quartet_1B_FSI_N2LOsim_575.dat}
CR_1S0 = 6.58
CR_3P0 = 0.58
CR_3P1 = 0.49
CR_3P2 = 1.44
CR_1D2 = 1.97
CR_3F2 = 0.27
CR_upto_J1 = 7.65
CR_sum = 11.33
\end{filecontents*}

\begin{filecontents*}{capRateListAvg_quartet_1B_FSI_N2LOsim_600.dat}
CR_1S0 = 6.57
CR_3P0 = 0.58
CR_3P1 = 0.49
CR_3P2 = 1.44
CR_1D2 = 1.98
CR_3F2 = 0.27
CR_upto_J1 = 7.64
CR_sum = 11.33
\end{filecontents*}

\begin{filecontents*}{capRateListAvg_doublet_1B2B_FSI_N2LOsim_450.dat}
CR_1S0 = 255.04
CR_3P0 = 15.27
CR_3P1 = 45.55
CR_3P2 = 71.88
CR_1D2 = 7.75
CR_3F2 = 0.97
CR_upto_J1 = 315.86
CR_sum = 396.46
\end{filecontents*}

\begin{filecontents*}{capRateListAvg_doublet_1B2B_FSI_N2LOsim_500.dat}
CR_1S0 = 254.44
CR_3P0 = 16.02
CR_3P1 = 45.72
CR_3P2 = 71.93
CR_1D2 = 7.75
CR_3F2 = 0.98
CR_upto_J1 = 316.18
CR_sum = 396.84
\end{filecontents*}

\begin{filecontents*}{capRateListAvg_doublet_1B2B_FSI_N2LOsim_550.dat}
CR_1S0 = 253.48
CR_3P0 = 16.32
CR_3P1 = 45.84
CR_3P2 = 72.00
CR_1D2 = 7.75
CR_3F2 = 0.98
CR_upto_J1 = 315.64
CR_sum = 396.37
\end{filecontents*}

\begin{filecontents*}{capRateListAvg_doublet_1B2B_FSI_N2LOsim_600.dat}
CR_1S0 = 252.09
CR_3P0 = 16.47
CR_3P1 = 45.89
CR_3P2 = 72.06
CR_1D2 = 7.76
CR_3F2 = 0.98
CR_upto_J1 = 314.45
CR_sum = 395.25
\end{filecontents*}

\begin{filecontents*}{capRateListAvg_quartet_1B2B_FSI_N2LOsim_450.dat}
CR_1S0 = 6.72
CR_3P0 = 0.54
CR_3P1 = 0.50
CR_3P2 = 1.53
CR_1D2 = 2.58
CR_3F2 = 0.34
CR_upto_J1 = 7.76
CR_sum = 12.21
\end{filecontents*}

\begin{filecontents*}{capRateListAvg_quartet_1B2B_FSI_N2LOsim_500.dat}
CR_1S0 = 6.71
CR_3P0 = 0.57
CR_3P1 = 0.51
CR_3P2 = 1.54
CR_1D2 = 2.68
CR_3F2 = 0.34
CR_upto_J1 = 7.79
CR_sum = 12.35
\end{filecontents*}

\begin{filecontents*}{capRateListAvg_quartet_1B2B_FSI_N2LOsim_550.dat}
CR_1S0 = 6.69
CR_3P0 = 0.58
CR_3P1 = 0.51
CR_3P2 = 1.56
CR_1D2 = 2.75
CR_3F2 = 0.35
CR_upto_J1 = 7.78
CR_sum = 12.44
\end{filecontents*}

\begin{filecontents*}{capRateListAvg_quartet_1B2B_FSI_N2LOsim_600.dat}
CR_1S0 = 6.68
CR_3P0 = 0.59
CR_3P1 = 0.51
CR_3P2 = 1.56
CR_1D2 = 2.79
CR_3F2 = 0.35
CR_upto_J1 = 7.78
CR_sum = 12.48
\end{filecontents*}

\begin{filecontents*}{capRateList_LO_doublet_1B2B_FSI_CHP_NNLO.dat}
CR_1S0 = 189.36
CR_3P0 = 13.60
CR_3P1 = 30.23
CR_3P2 = 58.98
CR_1D2 = 5.88
CR_3F2 = 0.70
CR_upto_J1 = 233.19
CR_sum = 298.75
\end{filecontents*}

\begin{filecontents*}{capRateList_NLO_doublet_1B2B_FSI_CHP_NNLO.dat}
CR_1S0 = 250.92
CR_3P0 = 19.28
CR_3P1 = 47.08
CR_3P2 = 72.05
CR_1D2 = 7.91
CR_3F2 = 1.00
CR_upto_J1 = 317.29
CR_sum = 398.24
\end{filecontents*}

\begin{filecontents*}{capRateList_N2LO_doublet_1B2B_FSI_CHP_NNLO.dat}
CR_1S0 = 254.39
CR_3P0 = 19.23
CR_3P1 = 47.07
CR_3P2 = 72.31
CR_1D2 = 7.89
CR_3F2 = 1.01
CR_upto_J1 = 320.69
CR_sum = 401.90
\end{filecontents*}

\begin{filecontents*}{capRateList_LO_quartet_1B2B_FSI_CHP_NNLO.dat}
CR_1S0 = 2.77
CR_3P0 = 0.40
CR_3P1 = 0.21
CR_3P2 = 0.64
CR_1D2 = 1.29
CR_3F2 = 0.13
CR_upto_J1 = 3.38
CR_sum = 5.44
\end{filecontents*}

\begin{filecontents*}{capRateList_NLO_quartet_1B2B_FSI_CHP_NNLO.dat}
CR_1S0 = 6.63
CR_3P0 = 0.75
CR_3P1 = 0.52
CR_3P2 = 1.48
CR_1D2 = 1.84
CR_3F2 = 0.33
CR_upto_J1 = 7.90
CR_sum = 11.55
\end{filecontents*}

\begin{filecontents*}{capRateList_N2LO_quartet_1B2B_FSI_CHP_NNLO.dat}
CR_1S0 = 6.72
CR_3P0 = 0.75
CR_3P1 = 0.51
CR_3P2 = 1.53
CR_1D2 = 2.39
CR_3F2 = 0.34
CR_upto_J1 = 7.98
CR_sum = 12.24
\end{filecontents*}

\title{Muon capture on the deuteron in chiral effective field theory}

\author{%
Jose Bonilla}
\email{jbonilla@vols.utk.edu}
\affiliation{Department of Physics and Astronomy, University of
  Tennessee, Knoxville, TN 37996, USA}

\author{%
Bijaya Acharya}
\email{bid@ornl.gov}
  \affiliation{Physics Division, Oak Ridge National Laboratory, Oak
  Ridge, TN 37831, USA}

\author{%
Lucas Platter}
\email{lplatter@utk.edu}
\affiliation{Department of Physics and Astronomy, University of
  Tennessee, Knoxville, TN 37996, USA}
\affiliation{Physics Division, Oak Ridge National Laboratory, Oak
  Ridge, TN 37831, USA}
\affiliation{Institut f\"ur Kernphysik, Technische Universit\"at Darmstadt, 64289 Darmstadt, Germany}
\affiliation{ExtreMe Matter Institute EMMI,
  GSI Helmholtzzentrum
für Schwerionenforschung GmbH, 64291 Darmstadt, Germany}

\begin{abstract}
  We consider the capture of a muon on a deuteron. An uncertainty
  analysis of the dominant channels is important for a careful
  analysis of forthcoming experimental data.  We quantify the
  theoretical uncertainties of chiral effective-field-theory
  predictions of the muon-deuteron capture rate from the relevant
  neutron-neutron partial wave channels in the final state.  We study
  the dependence on the cutoff used to regularize the interactions,
  low-energy constants calibrated using different fitting data and
  strategies, and truncation of the effective-field-theory expansion
  of the currents. Combining these approaches gives as an estimate of
  $\Gamma^{1/2}_{\mu d} = 399.1 \pm 7.6 \pm 4.4$ s$^{-1}$ for capture
  from the atomic doublet state, and
  $\Gamma^{3/2}_{\mu d} = 12.31 \pm 0.47 \pm 0.04$ s$^{-1}$ for capture
  from the quartet state.
\end{abstract}
 \date{\today}
\smallskip
\maketitle
\newpage
\section{Introduction}
\label{sec:intro}
One of the main current priorities of nuclear theory is the description of 
nuclear electroweak processes. They give insights into the structure of complex
nuclei, can be used to search for physics beyond the standard model, and 
are also important inputs to models of big bang nucleosynthesis
and stellar evolution. Their calculation requires models of the
nuclear interaction and electroweak currents that are consistent
with each other. Chiral effective field theory provides a systematic approach
to derive these consistently within one
framework~\cite{Bedaque:2002mn,Epelbaum:2008ga,
  Hammer:2012id}.

  Effective field theories (EFTs) are systematic
low-energy expansions that can be constructed when a system displays a
separation of scales whose ratios can be used as the expansion
parameters. Within chiral EFT, nucleons and pions are the degrees of
freedom used to construct the nuclear Hamiltonian. The expansion
parameter $Q$ of chiral EFT is given by the ratio of the pion mass or a typical low momentum 
scale relevant for the problem at hand to $\Lambda_b$, the breakdown
scale of the theory, which is expected to be comparable to the
lightest degree of freedom not taken into account in the theory. 
The cost of this simplified EFT description of low-energy dynamics are
additional parameters in the EFT, known as low-energy constants
(LECs), that have to be determined by fitting to experiment or to
calculations with the underlying theory. One important example of such
parameters in chiral EFT are the two coupling constants $c_D$ and
$c_E$ of the leading chiral three-body force whose values have to be
determined by {\it matching} a theoretical calculation to experimental
data. One of the two parameters is only related to short-distance
three-nucleon physics, while the other is also related to the coupling
of the electroweak current to the two-nucleon system. It should
therefore be possible to obtain this coupling constant from an
experimental measurement that involves only two nucleons. Muon capture
on the deuteron, {\it i.e.} the process
$\mu^- + d \rightarrow \nu_\mu + n + n$, is one such process that is
experimentally accessible. The current operator thus calibrated can
then be used to make predictions for other nuclear electroweak
observables, {\it e.g.} the proton-proton fusion rate that serves as
important input to astrophysical models but can not be measured at
relevant energies.

Muon capture on nuclei has been a tool to study nuclear physics for a
long time and the rate of muon capture on the deuteron has been
experimentally measured several times in the
past~\cite{Wang:1965zzb,Bertin:1973xdh,Martino:1986gq,Cargnelli:1989}.
The precision of existing data, however, is not sufficient to guide
theoretical studies. An ongoing experiment at the Paul Scherrer
Institute aims to measure this rate with 1.5\%
precision~\cite{Kammel:2021jss}.  This will provide a strong
constraint on the two-nucleon axial current operator which will be
completely independent of the many-body dynamics that affect the
extraction of $c_D$ from $A\geq3$ observables.

On the theoretical side, this process has been considered previously
using different approaches, see for example Ref. \cite{Measday:2001yr}
and references therein. The first chiral EFT calculation of muon
capture into the neutron-neutron ($nn$) singlet $S$-wave was carried
out by Ando {\it et al. }\cite{Ando:2001es}. More recently, more
complete calculations of this rate were carried out in
Refs.~\cite{Marcucci:2011jm,Marcucci:2010ts,Golak:2016zcw}. In
Ref.~\cite{PhysRevC.98.065506}, some of us considered previously the
capture rate in chiral EFT with a focus on the $^1S_0$ neutron-neutron
final state channel. This channel gives the dominant contribution to
the capture rate and is the only channel that is sensitive to the
leading two-nucleon axial current in the chiral EFT expansion.  In
this work, we use the same chiral EFT interactions to consistently
include higher partial-wave contributions, which is necessary to
relate the $^1S_0$ capture rate to the experimental datum.  This
manuscript is ordered as follows. We discuss the electroweak current
in chiral EFT in Section~\ref{sec:currents} and summarize the
theoretical derivations needed to evaluate the muon capture rate in
Section~\ref{sec:calc-capt-rate}. We then present our findings and put
them in the context of previous literature in
Section~\ref{sec:capture_rate}.  We conclude with a brief summary and
outlook in Section~\ref{sec:conclusion}.

\section{\bf Electroweak currents}
\label{sec:currents}
Interactions between a system of particles and external sources are
described by current operators that allow the transition from an
initial state to a final state. In our case, these operators are the
building blocks of the nuclear electroweak current $J^\mu$ that is
written as a sum of vector and axial currents $V^\mu$ and $A^\mu$,
respectively
\begin{equation}
	J^{\mu} = A^{\mu}_{1B} + V^{\mu}_{1B} + A^{\mu}_{2B} + V^{\mu}_{2B}~,
\end{equation}
where the subscripts $1B$ and $2B$ indicate whether we are considering
a one-nucleon or two-nucleon current, respectively. Expressions for these
currents were previously derived in
Refs.~\cite{Park:1995pn,Park:1998wq,Park:2002yp,Song:2008zf}. 
In this work, we use the currents derived with the method of
unitary transformations by K\"olling {\it et
  al.}~\cite{Kolling:2011mt} and Krebs {\it et
  al.}~\cite{Krebs:2016rqz}. 
Consistent with the truncation of the nuclear potentials employed in
the computation of the wavefunctions, we take into account current
operators derived up to $Q^0$ (NNLO). Higher order terms are
suppressed but the theory uncertainty from neglecting them are, as we
will show below, comparable to the expected experimental
uncertainties.

The current operators used in our work are displayed in
Tbl.~\ref{tab:Q-ordering}, in which there are a variety of
non-vanishing leading order (LO), next-to-leading order (NLO), and
next-to-next-to-leading order (NNLO) contributions. The power-counting also includes
relativistic corrections that are denoted by terms that have $1/m$ as
subscript. Detailed expressions for the different terms for both axial
and vector currents are given in
Appendix~\ref{sec:electroweak-currents}.

\begin{table}[t]
	\begin{tabular*}{1.0\textwidth}{@{\extracolsep{\fill}} l |c c c}
		$J^{\mu}$ & $Q^{-3}$ (LO) & $Q^{-1}$ (NLO) & $Q^0$ (NNLO) \\
		\hline
		$A^0$& - & $A^0_{1B:UT} + A^0_{1B:1/m} + A^0_{2B:1\pi}$ & - \\
		$A^ i$& $A^i_{1B:static}$  & - & $A^i_{2B:1\pi} + A^i_{2B:cont}$ \\
		$ V^0$& $V^0_{1B:static}$ & - & - \\
		$ V^i$ & - & $V^i_{1B:static} + V^i_{1B:1/m} + V^i_{2B:1\pi}$ & - \\
	\end{tabular*}
      \caption{Ordering of the chiral electroweak currents as
        discussed in Refs.~\cite{Kolling:2011mt,Krebs:2016rqz}. Terms
        with a subscript $static$ denote the contributions in which
        the external current couples directly to the nucleon, a
        subscript $1/m$ denotes relativistic corrections and the
        subscript $1\pi$ denotes contributions that include a pion
        loop.}
	\label{tab:Q-ordering}
\end{table}

The first contribution to the total electroweak current appears at
order $Q^{-3}$ that includes a static one-body time-like vector operator
Eq.~\eqref{eq:vector_0_1B} and a one-body space-like axial operator
Eq.~\eqref{eq:axial_i_1B} which consists of the sum of the well known
Gamow-Teller operator and a pion-pole contribution that is contained
in the pseudoscalar form factor of this term. At order $Q^{-1}$, we
encounter the one-body time-like axial operator
Eq.~\eqref{eq:axial_0_1B} which emerges from the time-dependence of
unitary transformations and a leading relativistic $1/m$
correction. Moreover we have a space-like vector current contribution,
shown in Eq.~\eqref{eq:vector_i_1B}, that includes the so-called
convection current and the spin-magnetization terms. At this order and
at $Q^0$, we include the two-body axial and vector current operators
Eqs.~\eqref{eq:axial_0_2B}, \eqref{eq:axial_i_2B},
\eqref{eq:axial_cont_i_2B}, and \eqref{eq:vector_i_2B}. We note that
the space-like axial operators of Eqs.  \eqref{eq:axial_i_2B} and
\eqref{eq:axial_cont_i_2B} feature LECs that also parametrize the
pion-nucleon and three-nucleon forces. These are represented by $c_i$
and $c_D$, respectively.
\section{Calculation of the capture rate}
\label{sec:calc-capt-rate}
To obtain the capture rate, we first calculate the momentum-space
matrix elements of the current operators discussed above, which are
needed to evaluate the corresponding transition amplitude, defined as
\begin{equation}
	\label{eq:MErate}
	T_{fi} = \frac{G_V}{\sqrt{2}} \psi_{\mu d}(0) \sum_{s_\mu, M_d}
	C^{f,f_z}_{1/2,s_\mu; 1 M_d }\ l^\sigma
	{}^{-}\braket{\Psi_f(\boldsymbol{p}) s_1 s_2| \hat{J}_\sigma| \psi_d M_d}~,
\end{equation}
where the incoming state $\ket{\psi_d M_d}$ is the deuteron bound
state with wave function $\psi_d$ and total angular momentum
projection $M_d$, and the outgoing state
$\ket{\Psi_f(\boldsymbol{p}) s_1 s_2}$ is a neutron-neutron scattering
state with wave function $\Psi_f$ and with spin projections $s_1$ and
$s_2$.  Using a complete set of momentum states, we write the deuteron
state as
\begin{equation}
	\ket{\psi_{d},M_d} = \sum_{l_d=0,2}\int dp p^2\ket{p(l_d1);1,M_d}\otimes\ket{0,0}\psi_{l_d}(p)~,
\end{equation}
and we express the scattering state with relative momentum ${\bf p}$
by using the identity
 \begin{equation}
  \bra{\Psi_f({\bf p}) s_1 s_2} = \bra {{\bf p} s_1 s_2 }\left[\hat{1} + \hat{t}(E_{nn}) G_0(E_{nn})\right]~,
\end{equation}
here $\hat{t}$ denotes the solution of the Lippmann-Schwinger equation
and $G_0$ is the free two-neutron Green's function, both evaluated at
the two-neutron scattering energy $E_{nn} = \frac{p^2}{m_n}$. The
leptonic tensor in Eq.~\eqref{eq:MErate} is given by
\begin{equation}
  l^\sigma = \bar{u}(k',h)\gamma^\sigma (1-\gamma_5) u(k,s_\mu)~,
\end{equation}
with lepton spinors $u(k,h)$. In addition to this, we employ a
coupling between the muon-deuteron spin by introducing a
Clebsch-Gordan coefficient $C^{f,f_z}_{1/2,s_\mu; 1 M_d}$ in
Eq.~\eqref{eq:MErate} which allows to calculate the capture rates for
the two hyperfine states $f=1/2$ and $f=3/2$.

Calculating the capture rate requires integration over the solid angle
of ${\bf p}$. To relate the capture rate to the matrix elements with
partial-wave projected final states, we express the transition
amplitude of Eq.~\eqref{eq:MErate} in terms of spherical harmonics.
\begin{multline}
	\label{eq:expandedME}
	T_{fi}  = \frac{G}{\sqrt{2}}\psi_{\mu d}(0) \sum_{s_\mu, M_d} C^{f, f_z}_{1/2, s_\mu; 1, M_d} l^\sigma
	\sum_{\alpha, m_l, s_z} Y^{*m_l}_{l}(\hat{p}) C^{J,M_J}_{l,m_l;s,s_z} C^{s, s_z}_{1/2, s_1; 1/2, s_2}
	\bra{ p \alpha}(\hat{1}+\hat{t} G_0) j_\sigma  \ket{\psi_d M_d}~.
\end{multline}
Here, $\alpha$ denotes the channel with quantum numbers
$\alpha \equiv \{(l s);J M_J\}$. In this work, we calculate the rate
up to a $J\leq2$ which includes the channels
$^1S_0$, $^3P_0$, $^3P_1$, $^3P_2$, $^1D_2$, $^3F_2$ that make
non-negligible contributions to the total muon
capture rate. The integral over the solid angle of ${\bf p}$ then
gives angle-averaged squared matrix elements which can be easily
related to the total capture rate.
\begin{equation}
	\Gamma^{f}_{\mu d} \propto \overline{|T_{fi}|^2} = \int d \hat{p}\frac{1}{2f+1}\sum_{f_z}\sum_{s_1,s_2}|T_{fi}|^2~.
\end{equation}

To obtain the unpolarized rate, we then sum over the spin projections  $s_1$ and $s_2$ of
the outgoing nucleons, leading to
\begin{multline}
	\overline{|T_{fi}|^2}=\frac{G_V^2}{2}|\psi_{\mu d}(0)|^2 \frac{1}{2f+1} \sum_{f_z}
	\sum_{\alpha}\Biggr| \sum_{s_\mu, M_d} C^{f, f_z}_{1/2, s_\mu; 1, M_d}\ l^\sigma
	\bra{ p \alpha}(\hat{1} + \hat{t} G_0) \hat{J}_\sigma \ket{\psi_d M_d}\Biggr|^2~.
\end{multline}
where
$\psi_{\mu d}(0) = [\alpha_\mathrm{em}
m_{\mu}m_{d}/(m_{\mu}+m_{d})]^{3/2}/\pi^{1/2}$ is the ground state
wavefunction of the muonic-deuterium atom at the origin and
$\alpha_\mathrm{em}$ denotes the fine structure constant.  Finally, the momentum
distribution of the capture rate for any channel can be calculated by
carrying out the phase space integral over the momentum of the
outgoing neutrino, which
yields
\begin{equation}
	\label{eq:diffCapRate}
	\frac{d\Gamma^{f}_{\mu d}}{dp} = \frac{2}{\pi} E_{\nu}^2 p^2\frac{m_n}{E_{\nu}+2m_n} \overline{|T_{fi}|^2} ~.
\end{equation}
The energy of the neutrino in Eq.~(\ref{eq:diffCapRate}) is given by
$E_{\nu} = \frac{1}{2(m_{\mu}+m_d)}\big[(m_{\mu}+m_d)^2 - 4(m_n^2 +
p^2)\big]$, where $m_{\mu}$, $m_d$, and $m_n$ are the masses of the
muon, deuteron, and the neutron respectively. 
The total capture rate $\Gamma^{f}_{\mu d}$ can be
calculated by integrating Eq.~(\ref{eq:diffCapRate}) over the relative
momentum $p$ from $0$ to
$p_{max} = \big[\frac{(m_{\mu}+m_d)^2}{4} - m_{n}^2\big]^{1/2}$.

\section{Results}
\label{sec:capture_rate}

\begin{table}[t]
	\begin{center}
		\scalebox{0.9}{
	\begin{tabular}{c|cccccc}
		\hline \hline
		\multirow{2}{*}{$g_A=1.2754$}       & \multicolumn{6}{c}{Capture Rate 1B $\Gamma^{f}_{\mu d}(s^{-1})$}                                                                                                                                 \\ \cline{2-7} 
		& \multicolumn{3}{c|}{$f=1/2$}                                                                           & \multicolumn{3}{c}{$f=3/2$}                                                        \\ \hline
		NNLOsim                  & \multicolumn{1}{c|}{$^1S_0$} & \multicolumn{1}{c|}{$J\leq1$} & \multicolumn{1}{c|}{$J\leq2$} & \multicolumn{1}{c|}{$^1S_0$} & \multicolumn{1}{c|}{$J\leq1$} & $J\leq2$\\ \hline
		$\Lambda$ = 450 MeV                 & \multicolumn{1}{c|}{\potA{CR_1S0}}    & \multicolumn{1}{c|}{\potA{CR_upto_J1}}               & \multicolumn{1}{c|}{\potA{CR_sum}}               & \multicolumn{1}{c|}{\potH{CR_1S0}}    & \multicolumn{1}{c|}{\potH{CR_upto_J1}}               & \potH{CR_sum}   \\
		$\Lambda$ = 475 MeV                 & \multicolumn{1}{c|}{\potB{CR_1S0}}    & \multicolumn{1}{c|}{\potB{CR_upto_J1}}               & \multicolumn{1}{c|}{\potB{CR_sum}}               & \multicolumn{1}{c|}{\potI{CR_1S0}}    & \multicolumn{1}{c|}{\potI{CR_upto_J1}}               & \potI{CR_sum}   \\
		$\Lambda$ = 500 MeV                 & \multicolumn{1}{c|}{\potC{CR_1S0}}    & \multicolumn{1}{c|}{\potC{CR_upto_J1}}               & \multicolumn{1}{c|}{\potC{CR_sum}}               & \multicolumn{1}{c|}{\potJ{CR_1S0}}    & \multicolumn{1}{c|}{\potJ{CR_upto_J1}}               & \potJ{CR_sum}   \\
		$\Lambda$ = 525 MeV                 & \multicolumn{1}{c|}{\potD{CR_1S0}}    & \multicolumn{1}{c|}{\potD{CR_upto_J1}}               & \multicolumn{1}{c|}{\potD{CR_sum}}               & \multicolumn{1}{c|}{\potK{CR_1S0}}    & \multicolumn{1}{c|}{\potK{CR_upto_J1}}               & \potK{CR_sum}                    \\
		$\Lambda$ = 550 MeV                 & \multicolumn{1}{c|}{\potE{CR_1S0}}    & \multicolumn{1}{c|}{\potE{CR_upto_J1}}               & \multicolumn{1}{c|}{\potE{CR_sum}}               & \multicolumn{1}{c|}{\potL{CR_1S0}}    & \multicolumn{1}{c|}{\potL{CR_upto_J1}}               &  \potL{CR_sum}                     \\
		$\Lambda$ = 575 MeV                 & \multicolumn{1}{c|}{\potF{CR_1S0}}    & \multicolumn{1}{c|}{\potF{CR_upto_J1}}               & \multicolumn{1}{c|}{\potF{CR_sum}}               & \multicolumn{1}{c|}{\potM{CR_1S0}}    & \multicolumn{1}{c|}{\potM{CR_upto_J1}}               & \potM{CR_sum}                      \\
		$\Lambda$ = 600 MeV                 & \multicolumn{1}{c|}{\potG{CR_1S0}}    & \multicolumn{1}{c|}{\potG{CR_upto_J1}}               & \multicolumn{1}{c|}{\potG{CR_sum}}               & \multicolumn{1}{c|}{\potN{CR_1S0}}    & \multicolumn{1}{c|}{\potN{CR_upto_J1}}               & \potN{CR_sum}                      \\ \hline
		\multicolumn{1}{c|}{A. Elmeshneb~\cite{Elmeshneb:2015tqr}} & \multicolumn{1}{c|}{240.5}    & \multicolumn{1}{c|}{303.3}               & \multicolumn{1}{c|}{383.4}               & \multicolumn{1}{c|}{6.38}    & \multicolumn{1}{c|}{7.73}               & \multicolumn{1}{c}{11.31} \\ \hline
	\end{tabular}}
      \caption{\label{tab:caprate-1B} Results for the muon capture
        rate for the doublet ($f=1/2$) and quartet ($f=3/2$) channel
        obtained with the NNLOsim~\cite{Carlsson:2015vda}
        interactions and NNLO one-body currents only. The
        different rows give the results obtained with different
        momentum cutoff $\Lambda$. Each given value is the average
        over the 6 NNLOsim interactions with different
        $T_{max}^{lab}$ truncations at this cutoff. Different columnns
        give the result for the rate with channels included up to
        $J=0,1$, or 2. The last row shows the corresponding results
        given in Ref.~\cite{Elmeshneb:2015tqr} for comparison.}
	\end{center}
\end{table}

\begin{table}[h]
	\begin{center}
		\scalebox{0.9}{
			\begin{tabular}{c|ccccccc}
				\hline \hline
				\multicolumn{7}{c}{Capture Rate for Doublet State 1B+2B $\Gamma^{1/2}_{\mu d}(s^{-1})$ }                                                                                                                                    
				\\ \hline \hline
				NNLOsim
				& \multicolumn{1}{c|}{$^1S_0$} & \multicolumn{1}{c|}{$^3P_0$} & \multicolumn{1}{c|}{$^3P_1$} & \multicolumn{1}{c|}{$^3P_2$} & \multicolumn{1}{c|}{$^1D_2$} & \multicolumn{1}{c|}{$^3F_2$} & Total \\ \hline
				$\Lambda$ = 450 MeV      & \multicolumn{1}{c|}{\potAA{CR_1S0}}    & \multicolumn{1}{c|}{\potAA{CR_3P0}}    & \multicolumn{1}{c|}{\potAA{CR_3P1}}    & \multicolumn{1}{c|}{\potAA{CR_3P2}}    & \multicolumn{1}{c|}{\potAA{CR_1D2}}    & \multicolumn{1}{c|}{\potAA{CR_3F2}}    & \multicolumn{1}{c}{\potAA{CR_sum}}     \\
				$\Lambda$ = 500 MeV     & \multicolumn{1}{c|}{\potBB{CR_1S0}}    & \multicolumn{1}{c|}{\potBB{CR_3P0}}    & \multicolumn{1}{c|}{\potBB{CR_3P1}}    & \multicolumn{1}{c|}{\potBB{CR_3P2}}    & \multicolumn{1}{c|}{\potBB{CR_1D2}}    & \multicolumn{1}{c|}{\potBB{CR_3F2}}    & \multicolumn{1}{c}{\potBB{CR_sum}}     \\
				$\Lambda$ = 550 MeV     & \multicolumn{1}{c|}{\potCC{CR_1S0}}    & \multicolumn{1}{c|}{\potCC{CR_3P0}}    & \multicolumn{1}{c|}{\potCC{CR_3P1}}    & \multicolumn{1}{c|}{\potCC{CR_3P2}}    & \multicolumn{1}{c|}{\potCC{CR_1D2}}    & \multicolumn{1}{c|}{\potCC{CR_3F2}}    & \multicolumn{1}{c}{\potCC{CR_sum}}     \\	
				$\Lambda$ = 600 MeV     & \multicolumn{1}{c|}{\potDD{CR_1S0}}    & \multicolumn{1}{c|}{\potDD{CR_3P0}}    & \multicolumn{1}{c|}{\potDD{CR_3P1}}    & \multicolumn{1}{c|}{\potDD{CR_3P2}}    & \multicolumn{1}{c|}{\potDD{CR_1D2}}    & \multicolumn{1}{c|}{\potDD{CR_3F2}}    & \multicolumn{1}{c}{\potDD{CR_sum}}     
				\\ \hline
			NNLO$_\mathrm{RS450}$ ($\Lambda$ = 450 MeV)
				& \multicolumn{1}{c|}{$^1S_0$} & \multicolumn{1}{c|}{$^3P_0$} & \multicolumn{1}{c|}{$^3P_1$} & \multicolumn{1}{c|}{$^3P_2$} & \multicolumn{1}{c|}{$^1D_2$} & \multicolumn{1}{c|}{$^3F_2$} & Total \\ \hline
				LO      & \multicolumn{1}{c|}{\potdLO{CR_1S0}}    & \multicolumn{1}{c|}{\potdLO{CR_3P0}}    & \multicolumn{1}{c|}{\potdLO{CR_3P1}}    & \multicolumn{1}{c|}{\potdLO{CR_3P2}}    & \multicolumn{1}{c|}{\potdLO{CR_1D2}}    & \multicolumn{1}{c|}{\potdLO{CR_3F2}}    & \multicolumn{1}{c}{\potdLO{CR_sum}}     \\
				NLO     & \multicolumn{1}{c|}{\potdNLO{CR_1S0}}    & \multicolumn{1}{c|}{\potdNLO{CR_3P0}}    & \multicolumn{1}{c|}{\potdNLO{CR_3P1}}    & \multicolumn{1}{c|}{\potdNLO{CR_3P2}}    & \multicolumn{1}{c|}{\potdNLO{CR_1D2}}    & \multicolumn{1}{c|}{\potdNLO{CR_3F2}}    & \multicolumn{1}{c}{\potdNLO{CR_sum}}    \\
				NNLO     & \multicolumn{1}{c|}{\potdNNLO{CR_1S0}}    & \multicolumn{1}{c|}{\potdNNLO{CR_3P0}}    & \multicolumn{1}{c|}{\potdNNLO{CR_3P1}}    & \multicolumn{1}{c|}{\potdNNLO{CR_3P2}}    & \multicolumn{1}{c|}{\potdNNLO{CR_1D2}}    & \multicolumn{1}{c|}{\potdNNLO{CR_3F2}}    & \multicolumn{1}{c}{\potdNNLO{CR_sum}}    
				\\ \hline
				Theoretical Results:\\
				\multicolumn{1}{c|}{S. Ando \emph{et al.} \cite{Ando:2001es}}  &  \multicolumn{2}{c}{386$\pm$4} \\
				\multicolumn{1}{c|}{L.E. Marcucci \emph{et al.} \cite{Marcucci:2011jm} }  &  \multicolumn{2}{c}{399$\pm$3} \\
				\multicolumn{1}{c|}{A. Elmeshneb~\cite{Elmeshneb:2015tqr} }  &  \multicolumn{2}{c}{401} \\ 	
				 \hline
				Experimental Results:\\
                          \multicolumn{1}{c|}{I.-T. Wang \emph{et al.} \cite{Wang:1965zzb}}  &  \multicolumn{2}{c}{365$\pm$96} \\
                          \multicolumn{1}{c|}{A. Bertin \emph{et al.} \cite{Bertin:1973xdh}}  &  \multicolumn{2}{c}{445$\pm$60} \\
                          \multicolumn{1}{c|}{M. Martino \cite{Martino:1986gq}}  &  \multicolumn{2}{c}{470$\pm$29} \\
                          \multicolumn{1}{c|}{M. Cargnelli \emph{et al.} \cite{Cargnelli:1989}}  &  \multicolumn{2}{c}{409$\pm$40} \\
                          \hline	\hline								
		\end{tabular}}
              \caption{Results for the muon capture rate for the
                doublet ($f=1/2$) channel in $s^{-1}$. Different columns give the
                results for the different partial wave channels
                included. The labels NNLOsim and
                NNLO$_{\rm RS}$ indicate the nucleon-nucleon
                interaction at order $Q^3$ used to calculate deuteron
                and $nn$ wave functions.}
		\label{tab:caprate-doublet-1B2B}
	\end{center}
\end{table}

\begin{table}[h]
	\begin{center}
		\scalebox{0.9}{
			\begin{tabular}{c|ccccccc}
				\hline \hline	
				\multicolumn{7}{c}{Capture Rate for Quartet State 1B+2B $\Gamma^{3/2}_{\mu d}(s^{-1})$ }                                                                                                                                    
				\\ \hline \hline
				NNLOsim
				& \multicolumn{1}{c|}{$^1S_0$} & \multicolumn{1}{c|}{$^3P_0$} & \multicolumn{1}{c|}{$^3P_1$} & \multicolumn{1}{c|}{$^3P_2$} & \multicolumn{1}{c|}{$^1D_2$} & \multicolumn{1}{c|}{$^3F_2$} & Total \\ \hline
				$\Lambda$ = 450 MeV      & \multicolumn{1}{c|}{\potEE{CR_1S0}}    & \multicolumn{1}{c|}{\potEE{CR_3P0}}    & \multicolumn{1}{c|}{\potEE{CR_3P1}}    & \multicolumn{1}{c|}{\potEE{CR_3P2}}    & \multicolumn{1}{c|}{\potEE{CR_1D2}}    & \multicolumn{1}{c|}{\potEE{CR_3F2}}    & \multicolumn{1}{c}{\potEE{CR_sum}}     \\
				$\Lambda$ = 500 MeV     & \multicolumn{1}{c|}{\potFF{CR_1S0}}    & \multicolumn{1}{c|}{\potFF{CR_3P0}}    & \multicolumn{1}{c|}{\potFF{CR_3P1}}    & \multicolumn{1}{c|}{\potFF{CR_3P2}}    & \multicolumn{1}{c|}{\potFF{CR_1D2}}    & \multicolumn{1}{c|}{\potFF{CR_3F2}}    & \multicolumn{1}{c}{\potFF{CR_sum}}     \\
				$\Lambda$ = 550 MeV     & \multicolumn{1}{c|}{\potGG{CR_1S0}}    & \multicolumn{1}{c|}{\potGG{CR_3P0}}    & \multicolumn{1}{c|}{\potGG{CR_3P1}}    & \multicolumn{1}{c|}{\potGG{CR_3P2}}    & \multicolumn{1}{c|}{\potGG{CR_1D2}}    & \multicolumn{1}{c|}{\potGG{CR_3F2}}    & \multicolumn{1}{c}{\potGG{CR_sum}}     \\	
				$\Lambda$ = 600 MeV     & \multicolumn{1}{c|}{\potHH{CR_1S0}}    & \multicolumn{1}{c|}{\potHH{CR_3P0}}    & \multicolumn{1}{c|}{\potHH{CR_3P1}}    & \multicolumn{1}{c|}{\potHH{CR_3P2}}    & \multicolumn{1}{c|}{\potHH{CR_1D2}}    & \multicolumn{1}{c|}{\potHH{CR_3F2}}    & \multicolumn{1}{c}{\potHH{CR_sum}}     \\ \hline			
				NNLO$_\mathrm{RS450}$ ($\Lambda$ = 450 MeV)
				& \multicolumn{1}{c|}{$^1S_0$} & \multicolumn{1}{c|}{$^3P_0$} & \multicolumn{1}{c|}{$^3P_1$} & \multicolumn{1}{c|}{$^3P_2$} & \multicolumn{1}{c|}{$^1D_2$} & \multicolumn{1}{c|}{$^3F_2$} & Total \\ \hline
				LO      & \multicolumn{1}{c|}{\potqLO{CR_1S0}}    & \multicolumn{1}{c|}{\potqLO{CR_3P0}}    & \multicolumn{1}{c|}{\potqLO{CR_3P1}}    & \multicolumn{1}{c|}{\potqLO{CR_3P2}}    & \multicolumn{1}{c|}{\potqLO{CR_1D2}}    & \multicolumn{1}{c|}{\potqLO{CR_3F2}}    & \multicolumn{1}{c}{\potqLO{CR_sum}}     \\
				NLO     & \multicolumn{1}{c|}{\potqNLO{CR_1S0}}    & \multicolumn{1}{c|}{\potqNLO{CR_3P0}}    & \multicolumn{1}{c|}{\potqNLO{CR_3P1}}    & \multicolumn{1}{c|}{\potqNLO{CR_3P2}}    & \multicolumn{1}{c|}{\potqNLO{CR_1D2}}    & \multicolumn{1}{c|}{\potqNLO{CR_3F2}}    & \multicolumn{1}{c}{\potqNLO{CR_sum}}    \\
				NNLO     & \multicolumn{1}{c|}{\potqNNLO{CR_1S0}}    & \multicolumn{1}{c|}{\potqNNLO{CR_3P0}}    & \multicolumn{1}{c|}{\potqNNLO{CR_3P1}}    & \multicolumn{1}{c|}{\potqNNLO{CR_3P2}}    & \multicolumn{1}{c|}{\potqNNLO{CR_1D2}}    & \multicolumn{1}{c|}{\potqNNLO{CR_3F2}}    & \multicolumn{1}{c}{\potqNNLO{CR_sum}}  
				\\ \hline
				Theoretical Results:\\
				\multicolumn{1}{c|}{A. Elmeshneb~\cite{Elmeshneb:2015tqr}}  &  \multicolumn{2}{c}{12.7}  
				\\ \hline \hline	
		\end{tabular}}
              \caption{Rate results for muon capture on the deuteron
                for hyperfine state $f=3/2$ (quartet). Results are
                calculated using partial wave decomposition approach
                and are displayed in each corresponding channel upto a
                $J\leq2$. The labels NNLOsim and NNLO$_{\rm RS}$
                indicate the nucleon-nucleon interaction at order
                $Q^3$ used to calculate deuteron and $nn$ wave
                functions. }
		\label{tab:caprate-quartet-1B2B}
	\end{center}
\end{table}

For the
calculation of the $\mu d$-capture rate, we first use a family of 42
interactions truncated at NNLO. The NN and NNN
LECs in these interactions have been fitted in
Ref.~\cite{Carlsson:2015vda} at seven different values of the
regulator cutoff $\Lambda$ at 25~MeV intervals in the range from 450
to 600 MeV simultaneously to the pion-nucleon data, the energies and
charge radii of $^{2,3}$H and $^3$He, the one-body quadrupole moment
of $^2$H, the comparative $\beta$-decay half life of $^3$H as well as
six different pools of NN scattering data with different truncations
in the NN scattering energy, $T_{\rm lab}$ .  These interactions,
which we denote by NNLOsim, have been refitted to account for a
correction~\cite{PhysRevLett.122.029901} in the equation that relates
$c_D$ to the axial two-body contact current (see
Ref.~\cite{PhysRevC.98.065506}), and then used to calculate muon
capture into the $^1$S$_0$
$nn$-channel~\cite{PhysRevC.98.065506}. Here, we have calculated the
rate for muon capture for the other five additional partial wave
channels that give a sizeable contribution to the rate and are
therefore important for comparison with experiment.  While the NNLOsim
interactions capture uncertainties from cutoff variation, sensitivity
to the input data sets, and fitting errors that account for
correlations among the LECs, it is also instructive to fix the
pion-nucleon LECs to the precise values obtained in
Refs.~\cite{Hoferichter:2015hva, Siemens:2017jb} using Roy-Steiner
analysis.  To this end, we use the LO, NLO and NNLO interactions of
Ref.~\cite{PhysRevC.104.064001}, which we name LO$_\mathrm{RS450}$,
NLO$_\mathrm{RS450}$, and NNLO$_\mathrm{RS450}$. These interactions
have been fit to the Granada
database~\cite{PhysRevC.88.064002,PhysRevC.91.054002,PhysRevC.95.064001}
as well as the $nn$ effective range parameters~\cite{Machleidt:2011zz}
with the pion-nucleon constants appearing at NNLO fixed at the central
values of the NLO~\footnote{Note that NLO in the pion-nucleon sector
  corresponds to NNLO in the NN interaction} pion-nucleon coupling
constants of Ref.~\cite{Siemens:2017jb}. The various orders of the
RS450 interactions also allow us to compare the uncertainty from the
truncation error in the potential to the NNLOsim uncertainties, which
is an important check of self-consistency of chiral
EFT~\cite{epelbaum202211}.

We show the results for the capture rates (capture from doublet and
quartet channel, and the channels included, truncated at different
total nuclear angular momentum $J_{\rm max}$) obtained with
differently regulated NNLOsim and one-body currents only in
Tbl.~\ref{tab:caprate-1B}. Here, each entry is the average over the
set of NNLOsim interactions with different $T_{\rm lab}$ truncations
at a given regulator $\Lambda$.  The last row shows the corresponding
results given in Ref.~\cite{Elmeshneb:2015tqr}. In
Tbl.~\ref{tab:caprate-doublet-1B2B}, we show the results for the
doublet capture rate obtained with NNLOsim potentials and one- and
two-body currents, the corresponding results obtained with the RS450
interactions at LO, NLO and NNLO, the experimental results given in
Refs.~\cite{Wang:1965zzb,Bertin:1973xdh,Martino:1986gq,Cargnelli:1989},
and the theoretical results obtained in
Refs.~\cite{Ando:2001es,Marcucci:2010ts, Elmeshneb:2015tqr}. The different columns give now the
contribution for each individual channel included with the last one
giving the total capture rate.

\begin{figure}[t]
  \includegraphics[width = 0.496\textwidth]{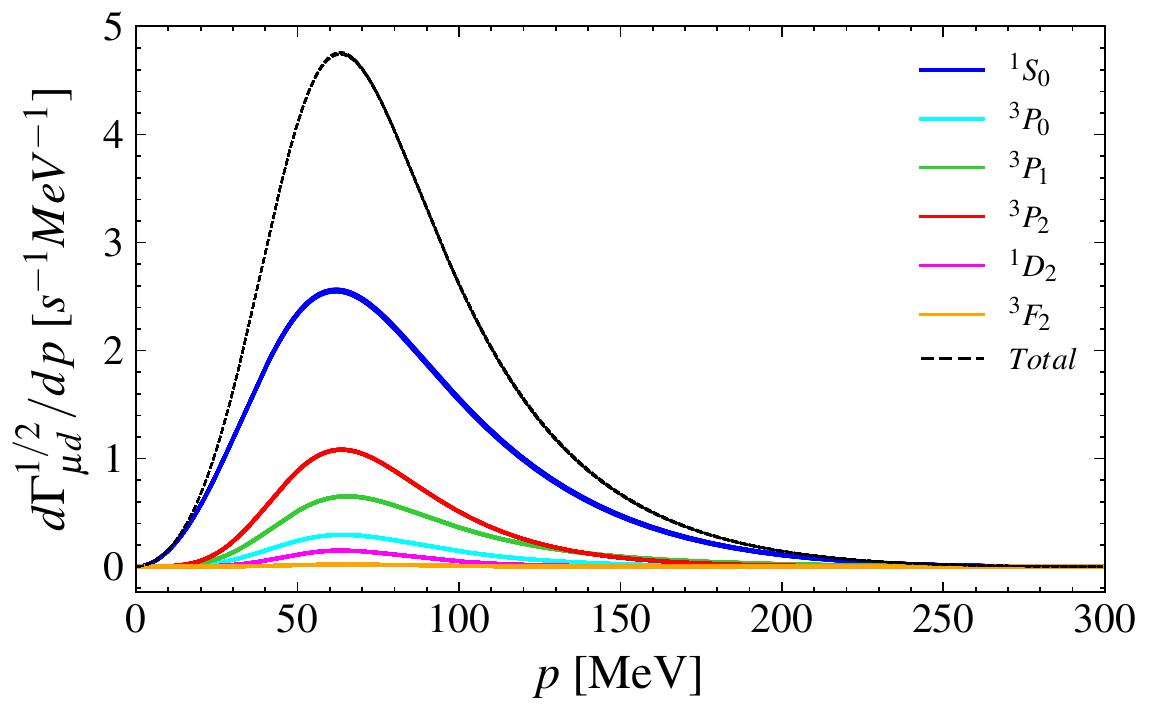}
  \includegraphics[width = 0.496 \textwidth]{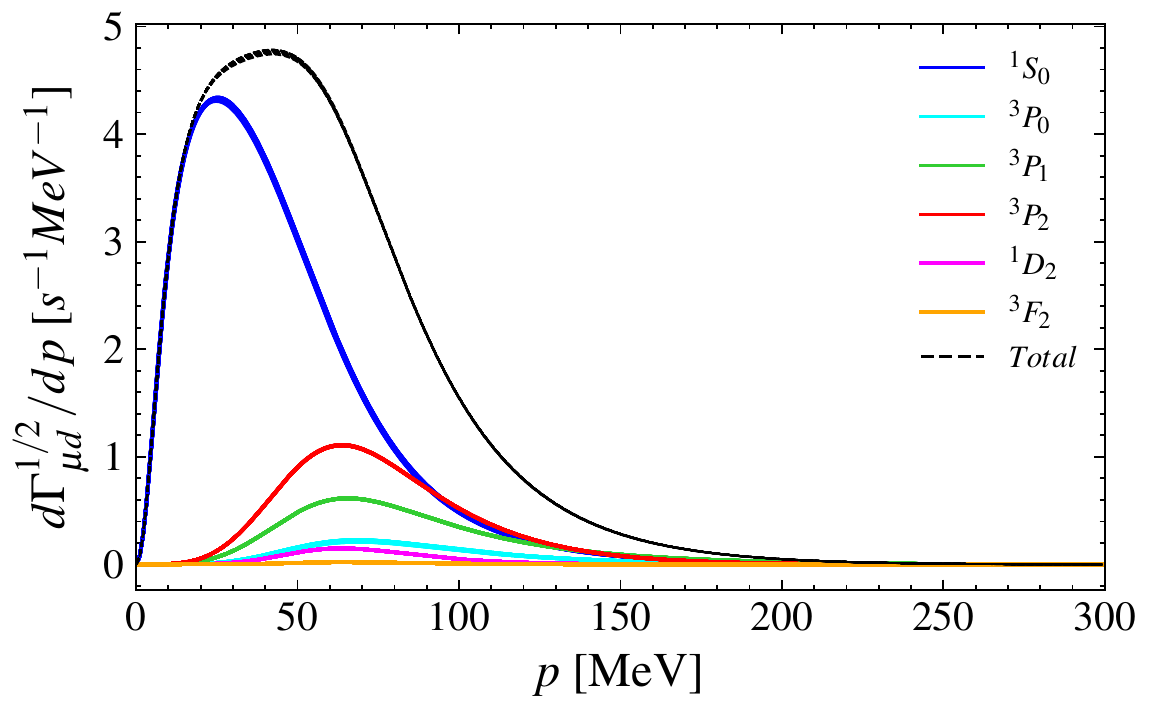}
  \caption{\label{fig:diff_cap_doublet} Left panel: Differential
    capture rate results for the doublet channel $f=1/2$ calculated
    without final state interactions. Right panel: Differential
    capture rate results for the doublet channel $f=1/2$ calculated
    with final state interactions. The solid lines give the results
    for different $nn$ partial wave channels. The dashed solid lines
    give the total differential capture rate.}
\end{figure}

\begin{figure}[t]
  \includegraphics[width = 0.496 \textwidth]{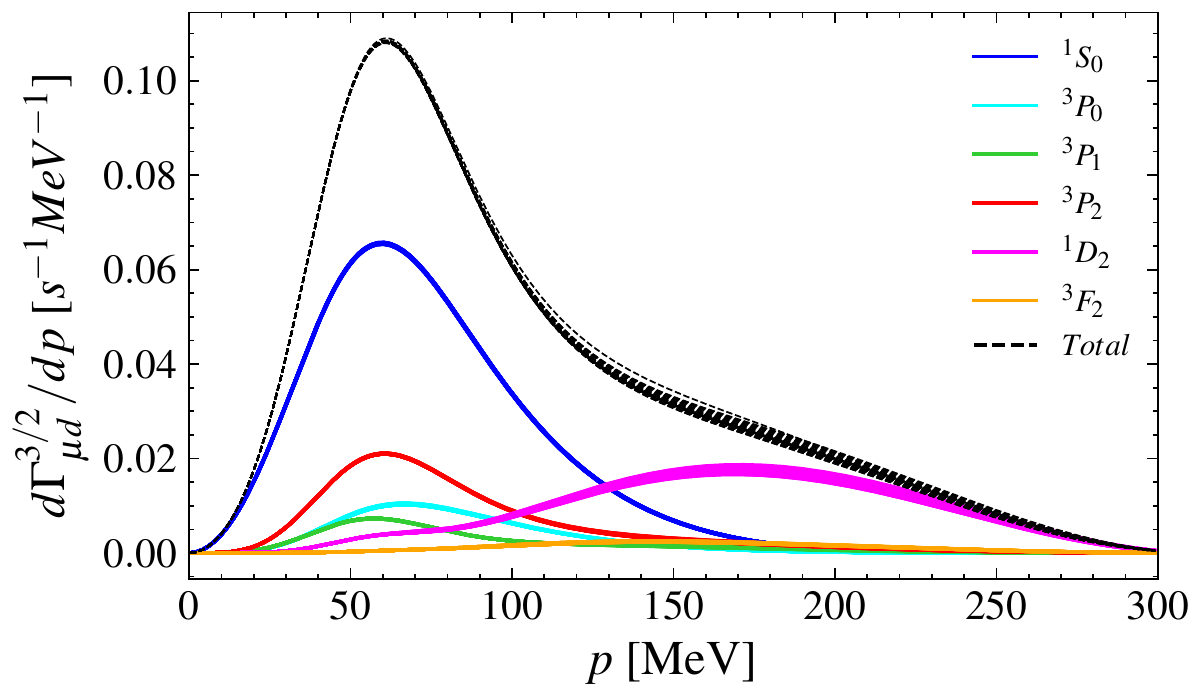}
  \includegraphics[width = 0.496 \textwidth]{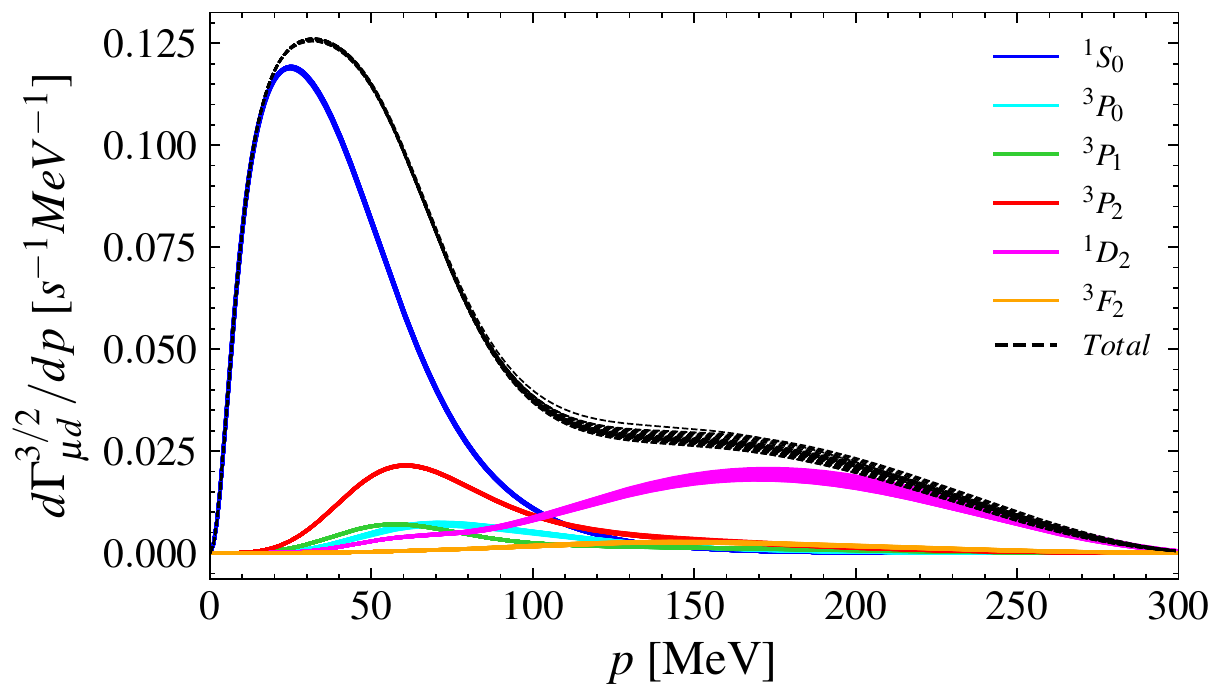}
  \caption{\label{fig:diff_cap_quartet} Left panel: Differential
    capture rate results for the quartet channel $f=3/2$ calculated
    without final state interactions. Right panel: Differential capture
    rate results for the quartet channel $f=3/2$ calculated with final
    state interactions. The solid lines give the results
    for different $nn$ partial wave channels. The dashed solid lines
    give the total differential capture rate.}
\end{figure}

To demonstrate the impact of the inclusion of final state
interactions, we show in Figs.~\ref{fig:diff_cap_doublet} and
\ref{fig:diff_cap_quartet}, the differential capture rate with
(right panel) and without final state interactions (left panel) as a
function of the magnitude of the relative momentum $p$ between the
outgoing neutrons for doublet and quartet channel, respectively. The
differently colored solid lines denote the contributions from the
individual partial wave channels of the $nn$ state. The dashed line
denotes the total differential capture rate. The widths of these lines
is generated through the calculation of the partial differential
capture rate with the different 42 different NNLOsim interactions. It
can clearly be seen that for both rates (doublet and quartet),
capture into the $^1S_0$ channel gives the largest contribution but also that a number
of different channels give sizeable contributions. The total
differential capture rates for doublet and quartet channel are in
qualitative agreement with the results shown in
Ref.~\cite{Elmeshneb:2015tqr} that were obtained with phenomenological
two-body currents.

\begin{figure}[t]
  \includegraphics[width = 0.496 \textwidth]{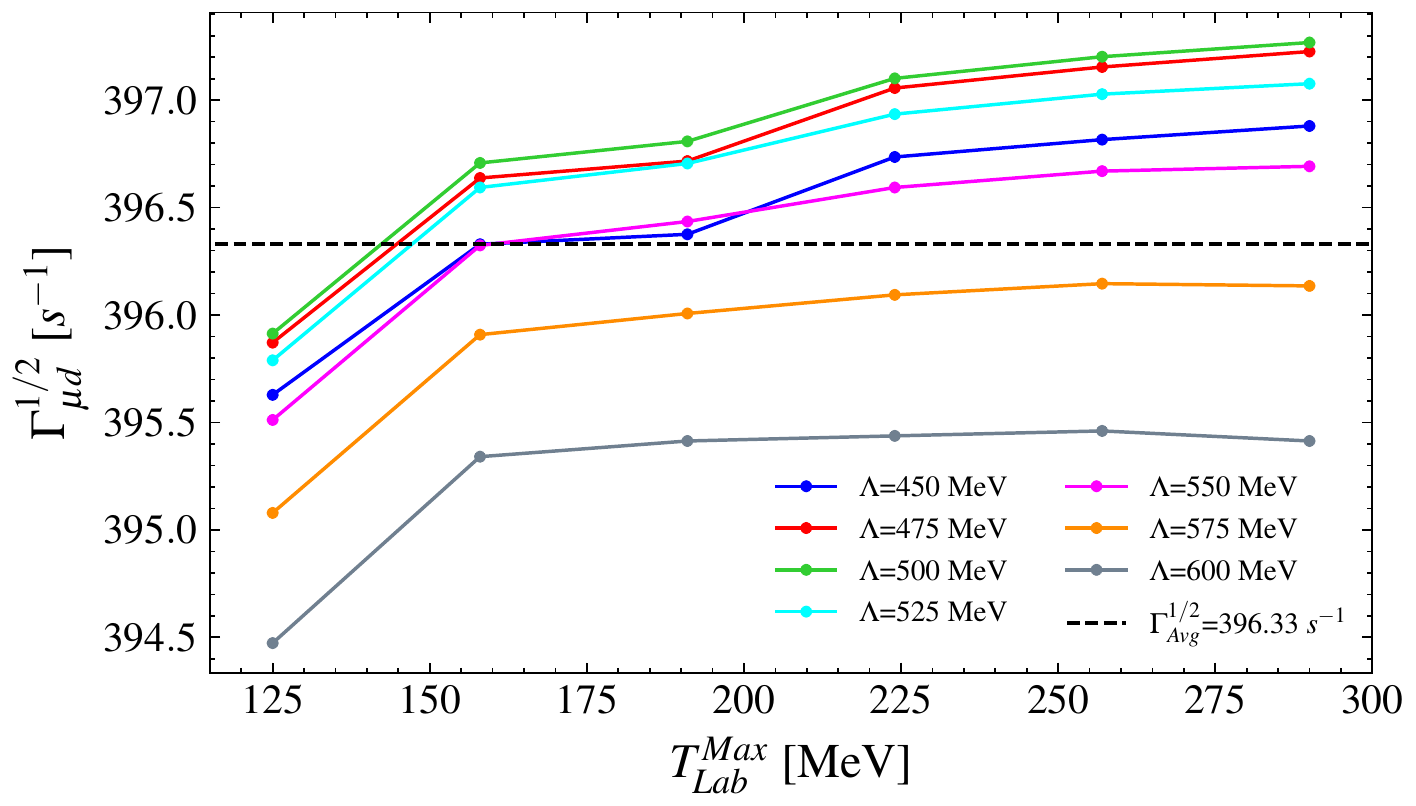}
  \includegraphics[width = 0.496 \textwidth]{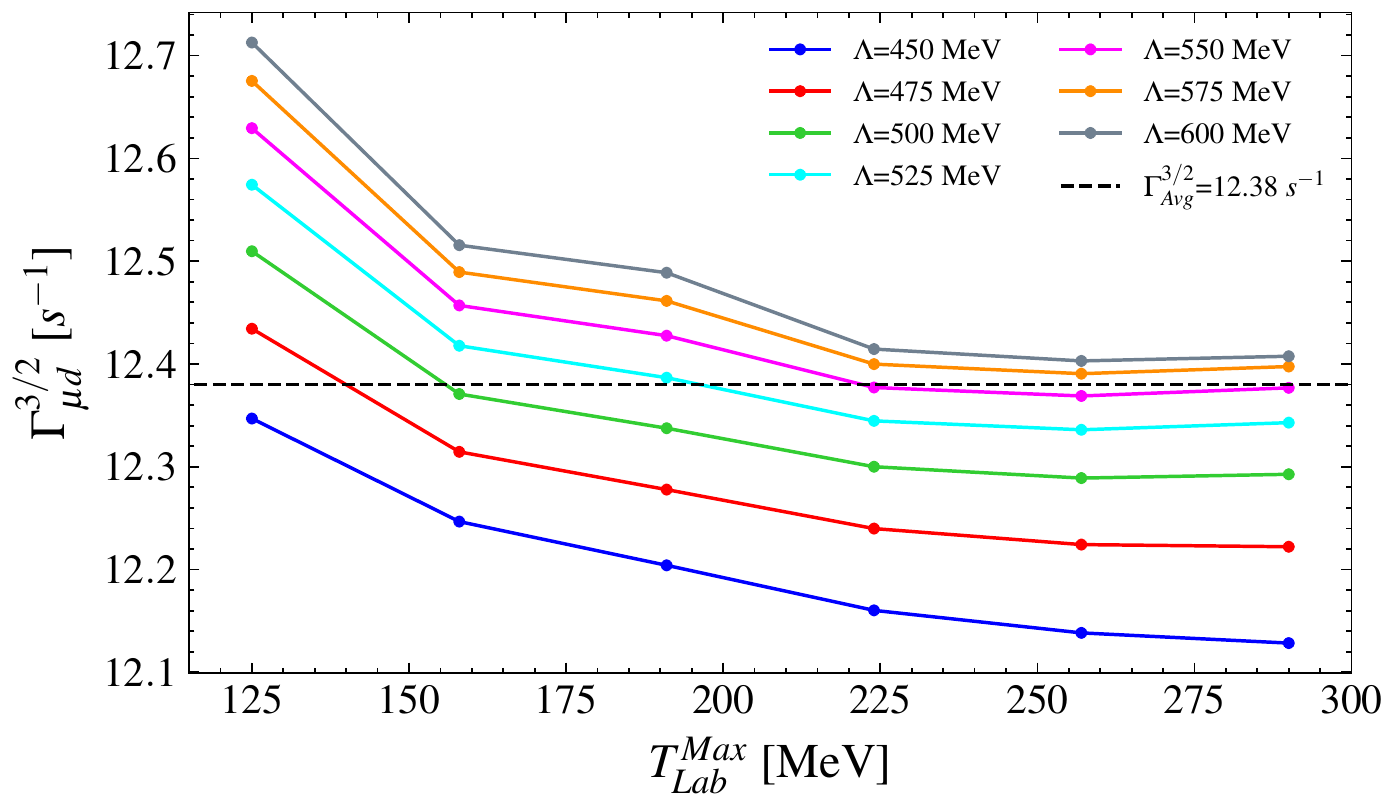}
        \caption{\label{fig:total_caprate} Left panel: Total muon capture rate
          on the deuteron evaluated for the
          doublet channel. Right panel: Total muon capture rate
          on the deuteron evaluated for the
          quartet channel. Results includes contributions for the $nn$-channels up to $J\leq2$. Each
          point represents the result obtained with one of the 42
          NNLOsim potentials at order $Q^3$ from
          Ref.~\cite{Carlsson:2015vda}. Results with the same cutoffs
          are connected by a line to guide the eye.}
\end{figure}

In the left (right) panel of Fig.~\ref{fig:total_caprate}, we show the
full rate $\Gamma^{1/2}_{\mu d}$ for capture from the doublet
(quartet) channel for the 42 different chiral interactions. We obtain
the central values of our rates by averaging the 42 results in each
channel. The spread between the smallest and largest rate and the
corresponding central value give us an estimate for the rate and its
uncertainty. This is shown as the first error in
Eq.~\eqref{eq:rate_NNLOsim_final} below.  We also propagate the
recently determined uncertainty in the axial radius
$r^2_A = 0.46\pm0.16$~fm$^2$~\cite{Hill:2017wgb} by calculating the
rates at the upper and lower range of this uncertainty estimate. This
is shown as the second (symmetrical) error in
Eq.~\eqref{eq:rate_NNLOsim_final} below. For the doublet and quartet
channel rate, we obtain in this way
\begin{align}
  \label{eq:rate_NNLOsim_final}
  \nonumber
\left[\Gamma^{1/2}_{\mu d}\right]_{\rm sim} &= (396.33^{0.94}_{-1.85} \pm 4.4) {\rm s}^{-1}~,\\
  \left[\Gamma^{3/2}_{\mu d}\right]_{\rm sim} &= (12.38^{0.34}_{-0.25} \pm 0.04) {\rm s}^{-1}~.
\end{align}
An even more reliable way to determine the uncertainty of an EFT
calculation is to study the order-by-order convergence pattern of an
observable. Here, we will follow the method discussed in
Ref.~\cite{PhysRevC.92.024005} by writing the capture rate for either doublet or quartet channel as
\begin{equation}
\Gamma_{\mu d} = \Gamma^{\rm LO}\sum_{n=0}^3 c_n\left( \frac{p}{\Lambda_b}\right)^n~,
\end{equation}
where $\Gamma_{\mu d}^{\rm LO}$ denotes the leading order result for
the muon capture rate (in either doublet or quartet channel), $p$
denotes the inherent momentum scale of the problem, and $\Lambda_b$ is
the breakdown scale. An estimate of the truncation is then obtained by
calculating $(p/\Lambda_b )^4 \max(|c_0 |, |c_2 |, |c_3 |)$. Using the
RS450 results of Tbl.~\ref{tab:caprate-doublet-1B2B} and Tbl.~\ref{tab:caprate-quartet-1B2B} to obtain the
$c_i$'s, the pion mass for the momentum scale $p$ and
$\Lambda_b = 500$~MeV, we obtain an uncertainty of 7.6~s$^{-1}$ for
the total doublet channel capture rate and 0.47~s$^{-1}$ for the total
quartet capture rate.

Using these results at face value we obtain for the $\mu d$-capture
from this approach gives
\begin{align}
  \label{eq:rate-RS450}
  \nonumber
  \left[\Gamma^{1/2}_{\mu d}\right]_{\rm RS450}  &= (401.90\pm 7.6 \pm 4.4)~{\rm s}^{-1}~,\\
  \left[\Gamma^{3/2}_{\mu d}\right]_{\rm RS450} & = (12.24 \pm 0.47 \pm 0.04)~{\rm s}^{-1}~,
\end{align}
where the first uncertainty quoted above is the estimate for the EFT
truncation error and the second is the uncertainty resulting from the
quoted uncertainty in the axial radius.  The uncertainty for the
doublet prediction is in good agreement with the uncertainty obtained
using the same method for capture into $^{1}$S$_0$ channel in
Ref.~\cite{PhysRevC.98.065506} but also with the spread in the final
results between NNLOsim and NNLO$_{\rm RS450}$ interactions.

To obtain a final recommendation for the $\mu d$-capture rates in
doublet an quartet channel we take the average of the values given in
Eqs.~\eqref{eq:rate_NNLOsim_final},\eqref{eq:rate-RS450} and use the
uncertainties of Eq.~\eqref{eq:rate-RS450}
\begin{align}
  \label{eq:rate-final}
  \Gamma^{1/2}_{\mu d}&= (399.1 \pm 7.6 \pm 4.4)~{\rm s}^{-1}~,\\
  \Gamma^{3/2}_{\mu d} & = (12.31 \pm 0.47 \pm 0.04)~{\rm s}^{-1}~.
\end{align}
Within the quoted truncation error, our results agree with the
previously published results in Refs.~\cite{Elmeshneb:2015tqr,
  Marcucci:2011jm} but disagrees slightly with the result given in
Ref.\cite{Ando:2001es}.

\section{ Conclusion}
\label{sec:conclusion}
In this work, we have calculated the total $\mu d$-capture from
doublet and quartet channel using chiral EFT potentials at NNLO and
consistent NNLO currents \cite{Kolling:2011mt,Krebs:2016rqz}. For the
total rates, we find after combining the results from different
interactions $\Gamma^{1/2}_{\mu d} = 399.1 \pm 7.6 \pm 4.4$ s$^{-1}$
for capture from the doublet channel, and
$\Gamma^{3/2}_{\mu d} = 12.31 \pm 0.47 \pm 0.04$ s$^{-1}$ for capture
from the quartet channel. The first uncertainty quoted above arises
from the order by order convergence pattern of the capture rates, the
second uncertainty propagates from the quoted uncertainty in the axial
radius. The recently determined large uncertainty of the axial radius
\cite{Hill:2017wgb} remains therefore a pressing problem for the
analysis of this problem as it prevents a reliable connection of the
experimental results with the nuclear Hamiltonian. However, the large
uncertainty in the doublet capture rate due to the intrinsic error of
chiral EFT also highlights that an experimental result for the capture
rate can provide important information on the nuclear Hamiltonian.

The results for the differential capture rate and the total capture
rate are in good agreement with previously published data
\cite{Ando:2001es,Marcucci:2011jm, Elmeshneb:2015tqr, Marcucci:2010ts}
apart from a small discrepancy with the $^1D_2$ doublet capture rate
quoted in \cite{Marcucci:2011jm}. To the best of our knowledge, this
work represents the first EFT calculation of the capture rate from the
quartet channel.

This and previous works demonstrate that muon capture on light nuclei
is a valuable tool to study the nuclear Hamiltonian. It is impacted by
superpositions of current matrix elements in a non-trivial way and
depends also strongly on an accurate descriptions of nuclear bound and
scattering properties. An ambitious program that focuses on the
reduction of experimental uncertainties and combined with capture rate
calculations for different processes might have the potential of
constraining the additional short-range counterterms that appear in
the axial and vector current at order
$Q$~\cite{Kolling:2011mt,Krebs:2016rqz}. Furthermore, a full
calculation of radiative corrections such as the two-photon exchange
contribution is desirable to assess their importance and their
dependence on nuclear structure effects.

\begin{acknowledgments}
  We acknowledge useful discussions with Evgeny Epelbaum, Jacek Golak
  and Andreas Ekstr\"om. This work has been supported by the National
  Science Foundation under Grant Nos. PHY-1555030 and PHY-2111426 and
  by the Office of Nuclear Physics, U.S.~Department of Energy under
  Contract No. DE-AC05-00OR22725. BA is supported by the Neutrino
  Theory Network Fellowship Program (Grant No. DE-AC02-07CH11359).
  This work used the Bridges-2 computing resource at the Pittsburgh
  Supercomputing Center through allocation PHY220101 from the Advanced
  Cyberinfrastructure Coordination Ecosystem: Services \& Support
  (ACCESS) program, which is supported by National Science Foundation
  grants \#2138259, \#2138286, \#2138307, \#2137603, and \#2138296.
\end{acknowledgments}

\begin{appendix}

\section{Electroweak Currents}
\label{sec:electroweak-currents}

\subsection{Axial currents}
The zero-component of the one-body axial current contains
contributions from the axial and pseudoscalar form factor
\begin{equation}
	\label{eq:axial_0_1B}
	A^{0}_{1B} = \bigg[-\frac{G_A(-{\bf q}^2)}{m}\q_1\cdotp \sa + \frac{G_P(-{\bf q}^2)}{4m^2}q_0\q\cdot\sa \bigg]\tau_{-,1}\ + \ (1 \rightarrow 2)~,
\end{equation}
where we use $m=938.9$~MeV and expressions for $G_A$ and $G_P$ will be
given below.  The vector components of the axial current are
\begin{equation}
	\label{eq:axial_i_1B}
	{\bf A}_{1B} = \bigg[-G_A(-{\bf q}^2)\sa + \frac{G_P(-{\bf q}^2)}{4m^2}\q(\q\cdot \sa)\bigg]\tau_{-,1} \ + \ (1 \rightarrow 2)~,
\end{equation}
where $\q = \p'_i-\p_i$, $\q_i = (\p'_i+\p_i)/2$, and $q_0 = (\p'^2_i-\p_i^2)/2m$.

The axial two-body currents used in this work have the form
\begin{equation}
	\label{eq:axial_0_2B}
	A^{0}_{2B:1\pi} = -i\frac{g_A}{4F^2_{\pi}}\frac{\ki_1\cdot\sa}{k_1^2+m_\pi^2}[\ta\times\tb]_{-} \ + \ (1 \leftrightarrow 2)~,
\end{equation}
where we use throughout this work $g_A = 1.2754$~\cite{Workman:2022ynf}, $F_{\pi} = 92.4$~MeV~\cite{Carlsson:2015vda}, and $m_{\pi} = 138.039$~MeV.
and
\begin{multline}
	\label{eq:axial_i_2B}
	{\bf A}_{2B:1\pi} = \frac{g_A}{2 F_\pi^2} \frac{\sa\cdot \ki_1}{k_1^2+m_\pi^2}
	\biggl\{\tau_{-,1}\left[-8c_1  m_\pi^2 \frac{\q}{q^2 + m_\pi^2}
	+4c_3\bigg(\ki_1 - \frac{\q\q\cdot \ki_1}{q^2+m_\pi^2}\bigg)\right]\\
	+c_4[\ta\times\tb]_{-}
	\bigg(\ki_1\times\sbb - \frac{\q\q\cdot \ki_1 \times \sbb}{q^2+m_\pi^2}\bigg)
	-\frac{\kappa_v}{4m}[\ta\times\tb]_{-} \q\times\sbb\biggr\} \ + \ (1 \leftrightarrow 2)
\end{multline}

\begin{equation}
 	\label{eq:axial_cont_i_2B}
        {\bf A}_{2B:cont}
        = -\frac{c_D}{2F_{\pi}^2 \Lambda_{\chi}}
        \bigg[\sa - \frac{\q(\sa\cdot\q)}{q^2 + \mpi^2}
        \bigg]
        \tau_{-,1} \ + \ (1 \leftrightarrow 2)~,
\end{equation}
where $\ki_i = \p'_i-\p_i$, $q = |\q|$, $\kappa_v$ is the isovector anomalous magnetic
moment of the nucleon,
$\Lambda_{\chi} = 700$~MeV is the chiral symmetry breaking scale of the order of
the $\rho$ meson mass.

\subsection{Vector currents}
The zero-component of the one-body vector current takes the standard
form
\begin{equation}
	\label{eq:vector_0_1B}
	V^{0}_{1B} = G_E(t) \ \tau_{-,1} \ + \ (1 \rightarrow 2)~.
\end{equation}
The spatial components of the one-body vector current operator receive the
standard contributions from the electric and magnetic couplings
encoded in the electric and magnetic form factors $G_E$ and $G_M$,
respectively
\begin{equation}
	\label{eq:vector_i_1B}
	{\bf V}_{1B} = \bigg[\frac{G_E(t)}{m}\q_1 -
        i\frac{G_M(t)}{2m}\big(\q\times\sa\big) \bigg] \tau_{-,1} \ + \ (1 \rightarrow 2)~,
\end{equation}
where we defined the four-momentum transfer
$ t = q_0^2 -{\bf q}^2 = m_{\mu}(m_{\mu} - 2E_{\nu})$, with ${\bf q} = E_{\nu}~\hat{z}$. The two-body
current that enters at NLO is
\begin{equation}
	\label{eq:vector_i_2B}
	{\bf V}_{2B:1\pi} = i\frac{g_A^2}{4F_{\pi}^2}\frac{\sbb\cdot\ki_2}{k_2^2 + \mpi^2}\bigg[\ki_1\frac{\sa\cdot\ki_1}{k_1^2 + \mpi^2} - \sa \bigg][\ta\times\tb]_{-}  \ + \ (1 \leftrightarrow 2)~.
\end{equation}

\subsection{Axial-Vector Form Factors}

\subsubsection{Axial Form Factors}
We parametrize the axial form factor as in Ref.~\cite{Krebs:2016rqz}
\begin{equation}
	G_A(-{\bf q}^2) = g_A\bigg(1 - \frac{\braket{r^2_A}}{6}{\bf q}^2\bigg)
\end{equation}
with the axial radius squared $\braket{r^2_A} = 0.46(16)$~fm$^2$.
The pseudoscalar form factor is 
\begin{equation}
	G_P(-{\bf q}^2) = \frac{4m^2}{{\bf q}^2+m_{\pi}^2}g_A~,
\end{equation}
where we emphasize that we found no difference in using this
parametrization versus using the form employed in
Ref.~\cite{Marcucci:2010ts} that replaces the factor of $g_A$ with the
axial form factor.
\subsubsection{Vector Form Factors}
In the currents defined above we employ the isovector combination of
the electric (magnetic) proton and neutron form factors $G_E^p$ and
$G_E^n$ ($G_M^p$ and $G_M^n$), respectively
\begin{equation}
	G_E = G^p_E - G^n_E \quad and \quad G_M = G^p_M - G^n_M
\end{equation}
The electric form factors are parametrized with dipole factors $G_D$      
\begin{equation}
	G^p_E(t) = G_D(t) \quad and \quad G^n_E(t) = \mu_n\frac{t}{4m^2}\frac{G_D(t)}{1-t/m^2}~,
\end{equation}
with the magnetic moments of proton and neutron $\mu_p = 2.793$
$\mu_n = -1.913$ in units of nuclear magnetons.
\begin{equation}
	G_D(t) = \frac{1}{(1-t/\Lambda_V^2)^2}~,
\end{equation}
where $\Lambda_V = 0.833$~GeV.
The magnetic form factors of proton and neutron are written as
\begin{equation}
	G^p_M(t) = \mu_p G_D(t) \quad and \quad G^n_M(t) = \mu_n G_D(t)~.
\end{equation}

\end{appendix}
\bibliographystyle{apsrev4-1}

%

\end{document}